\documentclass{article}
\usepackage{graphicx}

\usepackage{epsfig}
\usepackage{color}
\date{\today}

\usepackage{amsmath}
\usepackage{amssymb}
\usepackage[hang,nooneline,scriptsize]{subfigure}
\begin{document}

\title{AdS solitons with conformal scalar hair}
\author{{\large Yves Brihaye 
}$^{\ddagger}$,
{\large Betti Hartmann }$^{\dagger}$
and
{\large Sardor Tojiev }$^{\dagger}$
\\ \\
$^{\ddagger}${\small Physique-Math\'ematique, Universite de
Mons-Hainaut, 7000 Mons, Belgium}\\
$^{\dagger}${\small School of Engineering and Science, Jacobs University Bremen,
28759 Bremen, Germany}  }

\date{\today}
\setlength{\footnotesep}{0.5\footnotesep}
\newcommand{\dd}{\mbox{d}}
\newcommand{\tr}{\mbox{tr}}
\newcommand{\la}{\lambda}
\newcommand{\ka}{\kappa}
\newcommand{\f}{\phi}
\newcommand{\vf}{\varphi}
\newcommand{\F}{\Phi}
\newcommand{\al}{\alpha}
\newcommand{\ga}{\gamma}
\newcommand{\de}{\delta}
\newcommand{\si}{\sigma}
\newcommand{\bomega}{\mbox{\boldmath $\omega$}}
\newcommand{\bsi}{\mbox{\boldmath $\sigma$}}
\newcommand{\bchi}{\mbox{\boldmath $\chi$}}
\newcommand{\bal}{\mbox{\boldmath $\alpha$}}
\newcommand{\bpsi}{\mbox{\boldmath $\psi$}}
\newcommand{\brho}{\mbox{\boldmath $\varrho$}}
\newcommand{\beps}{\mbox{\boldmath $\varepsilon$}}
\newcommand{\bxi}{\mbox{\boldmath $\xi$}}
\newcommand{\bbeta}{\mbox{\boldmath $\beta$}}
\newcommand{\ee}{\end{equation}}
\newcommand{\eea}{\end{eqnarray}}
\newcommand{\be}{\begin{equation}}
\newcommand{\bea}{\begin{eqnarray}}

\newcommand{\ii}{\mbox{i}}
\newcommand{\e}{\mbox{e}}
\newcommand{\pa}{\partial}
\newcommand{\Om}{\Omega}
\newcommand{\vep}{\varepsilon}
\newcommand{\bfR}{{\bf R}}
\newcommand{\bfph}{{\bf \phi}}
\newcommand{\lm}{\lambda}
\def\theequation{\arabic{equation}}
\topmargin= -02cm\textheight= 23.cm\textwidth= 16.cm
\oddsidemargin=-01cm\evensidemargin=-01cm \font\sqi=cmssq8
\renewcommand{\thefootnote}{\fnsymbol{footnote}}
\newcommand{\re}[1]{(\ref{#1})}
\newcommand{\R}{{\rm I \hspace{-0.52ex} R}}
\newcommand{\N}{{\sf N\hspace*{-1.0ex}\rule{0.15ex}
{1.3ex}\hspace*{1.0ex}}}
\newcommand{\Q}{{\sf Q\hspace*{-1.1ex}\rule{0.15ex}
{1.5ex}\hspace*{1.1ex}}}
\newcommand{\C}{{\sf C\hspace*{-0.9ex}\rule{0.15ex}
{1.3ex}\hspace*{0.9ex}}}
\newcommand{\eins}{1\hspace{-0.56ex}{\rm I}}
\renewcommand{\thefootnote}{\arabic{footnote}}
 \maketitle

\begin{abstract}
We study  spherically symmetric soliton solutions in a model with a conformally coupled scalar field
as well as in full conformal gravity. We observe that a new type of limiting behaviour appears for particular choices of the
self-coupling of the scalar field, i.e. the solitons interpolate smoothly between the Anti-de Sitter vacuum and an uncharged configuration.
Furthermore, within conformal gravity the
qualitative approach of a limiting solution does not change when varying the charge of the scalar field
- contrary to the Einstein-Hilbert case. However, it changes with the scalar self-coupling. 
\end{abstract}


\section{Introduction}
The Anti-de Sitter (AdS)/Conformal Field theory (CFT) correspondence relates gravity theories
in $(d+1)$-dimensional asymptotically AdS space-time to a CFT living on the $d$-dimensional boundary of that
space-time \cite{ggdual,adscft}. As such, classical solutions in asymptotically AdS have gained
a lot of interest. This includes both black hole solutions as well as globally
regular, solitonic-like solutions. Within the context of the holographic description
of high-temperature superconductivity solutions with asymptotic planar AdS have been considered 
\cite{gubser,hhh,horowitz_roberts,reviews}. It was shown that the formation of scalar (or vector) ``hair''
on the solutions is the dual description of the onset of superconductivity. This is possible
since the effective mass of the scalar field drops below the Breitenlohner-Freedman (BF) bound \cite{bf}
close to the horizon if the black hole is close to extremality and hence the black hole becomes unstable to the 
formation of scalar hair on the horizon. Asymptotically, however, the BF bound is fullfilled and
hence the solutions tend to AdS space-time asymtotically. 
Besides being interesting due to the holographic interpretation, the stability of
classical field theory solutions is, of course, also of interest by itself as it could well shed light on
other interesting questions, e.g. black hole uniqueness. Consequently, the stability of black holes and solitons in global
AdS as well as in asymptotic hyperbolic AdS have also been discussed.
In \cite{Dias:2010ma} uncharged black holes in $(4+1)$ dimensions have been considered. It was shown that
static black holes with hyperbolic
horizons can become unstable to the formation of uncharged scalar hair on the horizon of the black hole
due to the existence of an extremal limit with near-horizon geometry AdS$_2\times H^3$ \cite{Robinson:1959ev,Bertotti:1959pf,Bardeen:1999px}. 
These studies were extended to include higher order curvature corrections in the form of 
Gauss-Bonnet terms \cite{hartmann_brihaye3}.

Charged black hole and soliton solutions in asymptotically global AdS in $(3+1)$ dimensions were studied in 
\cite{menagerie}.
It was pointed out  that the solutions tend to their planar counterparts for large charges since 
in that case the solutions can become comparable in size to the AdS radius. The influence of the Gauss-Bonnet
corrections on the instability of these solutions has been discussed in \cite{Brihaye:2012cb}. 
The corresponding investigation in $(4+1)$-dimensional global AdS was done in \cite{dias2,basu}. The existence
of solitons in this case had been suggested previously in a perturbative approach \cite{basu}. In all studies
the mass of the scalar field had been set to zero. In \cite{brihaye_hartmannNEW} these results were extended 
to a tachyonic scalar field as well as to the rotating case. Recently, charged soliton solutions with positive
scalar field mass have been studied in $(3+1)$-dimensional global AdS \cite{hu}. In this case, the solutions carrying scalar
hair can be interpreted as charged
non-spinning boson star solutions in global AdS space-time. Uncharged boson stars in AdS were first discussed in 
$(d+1)$-dimensional AdS space-time using a massive
scalar field without self-interaction \cite{radu} and with an exponential self-interaction
potential \cite{hartmann_riedel,hartmann_riedel2}, respectively. Spinning solutions in $(2+1)$ and $(3+1)$ dimensions
have been constructed in \cite{radu_aste} and \cite{radu_subagyo}, respectively. 

In most studies done so far, the scalar field is minimally coupled to Einstein-Hilbert gravity. One could however also consider 
a non-minimal coupling between gravity and the scalar field. This has been done for solitons and black holes in 
an AdS
background in \cite{Radu:2005bp}, while electrically charged AdS black holes with non-minimally coupled
scalar fields with planar horizon topology have recently been discussed \cite{Caldarelli:2013gqa}. 
One motivation for considering such a coupling comes from the observation
that the Higgs field that seems to have been discovered recently at the LHC \cite{cern_lhc} could play the role of the scalar field
responsible for inflation, the so-called inflaton, if this scalar field is non-minimally coupled to gravity \cite{Bezrukov:2007ep}.

On the gravity side essentially only Einstein-Hilbert gravity \cite{menagerie,dias2} or Einstein-Gauss-Bonnet gravity \cite{Brihaye:2012cb,bhs2}
has been considered in the context of solitons in AdS.
Here, we would like to study conformal gravity. This gravity theory has been proposed as an alternative to covariant
Einstein gravity and has the advantage that as soon as one requires additionally to covariance that the theory be locally
conformally invariant the action of the model is uniquely specified and is given by the Weyl action \cite{Mannheim2006}. As an example
let us mention that the cosmological constant has to be introduced ``by hand'' into the Einstein-Hilbert action, while
it naturally appears in Conformal Gravity. Furthermore, since the coupling constant appearing in conformal gravity
is dimensionless the theory becomes power-counting renormalizable in contrast to Einstein-Hilbert gravity which is plagued with the
dimensionful Newton's constant $G$.  

In this paper, we are interested in studying the formation of conformal scalar hair on
charged solitons in a $(3+1)$-dimensional asymptotically AdS space-time. While in previous studies the mass of the scalar field
was often chosen to be non-vanishing we choose it to be zero here and introduce a self-interaction
term into the matter Lagrangian that is invariant under conformal transformations. We will first be interested
to study the properties of the soliton solutions in the case where the matter Lagrangian is minimally
coupled to Einstein-Hilbert gravity. 
Then, we will construct solutions of a model in which
the scalar field is conformally coupled to Einstein-Hilbert gravity. Finally, we study a model whose
total Lagrangian is conformally invariant - including the gravity part. The gravity part will be given
by Weyl gravity. Spherically symmetric solutions in conformal gravity were studied in \cite{BrihayeVerbin}.

This paper is organized as follow: in Section 2, we present the general set-up, while
in section 3 and 4, respectively, we discuss the case of Einstein-Hilbert gravity and Weyl gravity
coupled to a scalar field.  Finally, Section 5 contains the conclusions.

\section{General set-up}

In this paper, we are studying the formation of scalar hair on electrically charged solitons
in asymptotically $(3+1)$-dimensional AdS space--time.
The action reads~:
\begin{equation}
S=S_{\rm gravity} + S_{\rm matter}= \int d^4 x \sqrt{-g} \left({\cal L}_{\rm gravity} +
{\cal L}_{\rm matter}\right) \ ,
\end{equation}
where $S_{\rm gravity}$ (${\cal L}_{\rm gravity}$) denotes the gravity action (Lagrangian) and 
$S_{\rm matter}$ (${\cal L}_{\rm matter}$) the matter action (Lagrangian), respectively and $g$ is the determinant
of the metric tensor $g_{\mu\nu}$. The matter Lagrangian reads~:
\begin{equation}
{\cal L}_{\rm matter}= -\frac{1}{4} F_{_{\mu\nu}} F^{^{\mu\nu}} -
\left(D_{_\mu}\psi\right)^* D^{^\mu} \psi - \beta R \vert\psi\vert^2 - V(\psi)  \ \ , \ \  
\mu, \nu=0,1,2,3        \ .
\end{equation}
$F_{_{\mu\nu}} =\partial_{_\mu} A_{_\nu} - \partial_{_\nu} A_{_\mu}$ is the field strength tensor of the U(1) gauge field
$A_{\mu}$ and
$D_{_\mu}\psi=\partial_{_\mu} \psi - ie A_{_\mu} \psi$ is the covariant derivative of the complex scalar field $\psi$ with
potential $V(\psi) =\lambda \vert\psi\vert^4$.
$e$ and $\lambda$ denote the electric charge and the self-coupling of the scalar field $\psi$, 
respectively.
Finally, $\beta$ is a parameter that can only have two values: either $\beta=0$ which corresponds to a
minimally coupled scalar field or $\beta=1/6$ which corresponds to a conformally coupled scalar field, respectively, with
$R$ the Ricci scalar. Note that the matter part of the action $S_{\rm matter}$ is invariant under a rescaling of the
type
\begin{equation}
 x^{\mu} \rightarrow \zeta x^{\mu} \  \ , \ \  A_{\mu} \rightarrow \zeta^{-1} A_{\mu} \ \ , \ \ \psi \rightarrow \zeta^{-1} \psi \ .
\end{equation}

In the following, we want to study static, spherically symmetric and electrically charged soliton solutions.
The Ansatz for the metric in Schwarzschild-like coordinates reads~:
\begin{equation}
\label{ansatz_metric}
ds^2 = g_{\mu\nu} dx^{\mu} dx^{\nu}=-a^2(r)f(r) dt^2 + \frac{1}{f(r)} dr^2 + r^2 \left(d\theta^2+\sin^2\theta d\varphi^2\right)  \ ,
\end{equation}
where the metric function $a(r)\equiv 1$ in the case of conformal gravity, while $a(r)\neq 1$ in general in the Einstein-Hilbert case.
For the electromagnetic field and the scalar field we choose~:
\begin{equation}
\label{ansatz_matter}
A_{_{\mu}}dx^{^\mu} = \phi(r) dt  \
\  , \   \   \   \psi=\psi(r)  \ .
\end{equation}
Note that originally the scalar field is complex, but that we can gauge away the non-trivial
phase and choose the scalar field to be real.

We will be interested in studying the dependence of the
physical properties of the soliton like the charge and the mass on the coupling constants. In the following, we will
use two different notions of the mass: the inertial mass $M_I$ and the gravitational mass $M_G$. The inertial mass is given by 
\cite{BrihayeVerbin} 
\be
     M_I =  \int d^3 x \sqrt{-g} T^0_0 = 4\pi \int\limits_0^{\infty} dr \ r^2 \ a(r) \ T^0_0(r)
\ee   
where the $T^{\mu}_{\nu}$, $\mu,\nu=0,1,2,3$ denote the components of the energy-momentum tensor. The gravitational 
mass $M_G$ has different definitions in Einstein-Hilbert gravity and conformal gravity, respectively. 
In Einstein-Hilbert gravity, the gravitational (or Komar) mass is defined by the
surface integral associated to the conserved charge related to time translation invariance.
It is proportional to the $1/r$ coefficient in the asymptotic behaviour of the metric function $f(r)$ 
in (\ref{ansatz_metric}). Using the field equations, it can further be expressed according to
\begin{equation}
 M_G=4\pi\int dr \ r^2 T^0_0(r)  \ .
\end{equation}
In conformal gravity we follow \cite{BrihayeVerbin} and define the gravitational mass as follows
\begin{equation}
 M_G=12\pi\int\limits_0^{\infty} dr \ r^2 \ \left(T^0_0(r) - T^r_r(r)\right)/f(r) \ .
\end{equation}
Obviously, even in Einstein-Hilbert gravity the ratio 
$M_G/M_I$ will not be constant since the integrand for $M_I$ contains an extra factor of the metric
function $a(r)$ which is not present for the gravitational mass. However, in the case of conformal gravity
we can use the conformal symmetry $g_{\mu\nu}\rightarrow \Omega g_{\mu\nu}$ to set the metric function
$a(r)\equiv 1$. Hence, the difference between the gravitational and inertial mass will indeed be a signal of the
violation of the equivalence principle in the case of conformal gravity.

In the following, we will be interested in two different type of gravity actions. First, we will study the coupling of the
matter field action to the standard Einstein-Hilbert action, where the latter is not conformally invariant.
Then we will discuss a matter-gravity action that as a whole is conformally invariant.

\section{Einstein-Hilbert gravity}
We would first like to study solutions in Einstein-Hilbert gravity in which case the gravity Lagrangian reads
\begin{equation}
 {\cal L}_{\rm gravity} = \frac{R - 2 \Lambda}{16\pi G} \ ,
\end{equation}
where the negative cosmological constant $\Lambda$ is related to the Anti-de Sitter radius $\ell$ by
$\Lambda=-3/\ell^2$ and $R$ is the Ricci scalar.

The coupled gravity and matter field equations are obtained from the variation of the
action with respect to the matter and metric fields, respectively, and read
\begin{equation}
 G_{_{\mu\nu}} + \Lambda g_{_{\mu\nu}}=8\pi G T_{_{\mu\nu}} \ ,  \ \mu, \nu=0,1,2,3 \ ,
\end{equation}
and
\begin{eqnarray}
D_\mu D^\mu \psi  + 2\lambda |\psi|^2 \psi + \beta R \psi = 0 \label{FieldEqsScalar} \ ,
\end{eqnarray}
\begin{eqnarray}
D_\mu F^{ \mu\nu} =-ie[\psi^* (D^\nu \psi)- (D^\nu \psi)^*  \psi ]  \ , 
\label{FieldEqsVector}
\end{eqnarray}
where $T_{_{\mu\nu}}$ is the energy-momentum tensor
\begin{equation}
{T}_{\mu\nu} = \tilde{T}_{\mu\nu} +{1\over 3}
\left (g_{\mu \nu} \nabla ^{\lambda }\nabla_
{\lambda } |\psi|^2 - \nabla _{\mu }\nabla_{\nu } |\psi|^2  - 
{G}_{\mu \nu} |\psi|^2 \right)
\label{confTmn}
\end{equation}
with $\tilde{T}_{\mu\nu}$ the standard ``minimal'' energy-momentum tensor of the matter fields given by
\begin{equation}
 \tilde{T}_{_{\mu\nu}}=F_{\mu\lambda}F^{\lambda}_{\nu}+D_{\mu}\psi (D_{\nu}\psi)^{*}+D_{\nu}\psi 
(D_{\mu}\psi)^{*}+g_{\mu\nu} {\cal L}_{\rm matter}  \ 
\end{equation}
and $G_{\mu\nu}$ is the Einstein tensor. 

Using the Ansatz (\ref{ansatz_metric}) and (\ref{ansatz_matter}) we then obtain a system of four coupled
ordinary differential equations, where the gravity equations for $f$ and $a$ are first order, while
the matter field equations for $\phi$ and $\psi$ are second order. The gravity equations are very lengthy that
is why we don't present them here. However, we give the matter field equations below to point out something
important related to the conformally coupled case. The matter equations read
\begin{eqnarray}
\label{eq3}
   \phi'' &=& -\left ( \frac{2}{r}- \frac{a'}{a} \right ) \phi' + \frac{2 e^2  \psi^2}{f} \phi \\
\label{eq4}
    \psi'' &=&-\left ( \frac{2}{r}+ \frac{f'}{f}+ \frac{a'}{a}  
\right ) \psi' - \frac{e^2  \phi^2 \psi}{a^2 f^2} + 2\lambda \frac{\psi^3}{f} + \beta \frac{R}{f}\psi
\end{eqnarray}
and the prime denotes
the derivative with respect to $r$. Note that if we would have included a mass term for the scalar field in the
action the equation for $\psi$ would contain a term of the form $m^2 \psi/f$. Since $R=-12/\ell^2$ asymptotically we find
that for $\beta=1/6$ the non-minimal coupling of the scalar field induces a mass term for the scalar field with $m^2=-2/\ell^2$.
The equations have  further scaling symmetries
\begin{equation}
 r\rightarrow \sigma r \ \ , \ \  t\rightarrow \sigma t \ \ , \ \ \ell \rightarrow \sigma \ell \ \ , \ \ e\rightarrow e/\sigma \ \ , \ \
\lambda \rightarrow \lambda/\sigma^2 \ ,
\end{equation}
\begin{equation}
 \psi\rightarrow \sigma \psi \ \ , \ \  \phi\rightarrow \sigma \phi \ \ , \ \ G \rightarrow G/\sigma^2 \ \ , \ \ e\rightarrow e/\sigma \ \ , \ \
\lambda \rightarrow \lambda/\sigma^2 \ .
\end{equation}
These can be used to set two of the coupling constants to a fixed value, in practice we will use this to set $\ell\equiv 1$ and $G$
to a fixed value. We will hence be left with two coupling constants that are too be varied: $\lambda$ and $e^2$.

In order to solve the system of coupled, non-linear differential equations we need to fix appropriate boundary conditions. 
The regularity of the solutions at $r=0$ requires
\begin{equation}
\phi'(0)=0 \ , \  \psi'(0)=0 \ , \ f'(0)=0 \ . 
\end{equation} 
On the AdS boundary, we choose $a(\infty)=1$, while the matter fields have the following behaviour
\be
       \psi(r>>1) = \frac{\psi_2}{r^2} + \frac{\psi_1}{r} \ \ \ , \ \ \ \phi(r>>1) = \mu - \frac{Q}{r} \ ,
\ee
where $Q$ corresponds to the charge of the soliton. In the following, we will fix $\psi_1\equiv 0$. Hence, we have the freedom to choose
one more parameter at will. In most case, we have chosen the value of $\psi(0)\equiv \psi_0$ which then fixes the values of $\psi_2$, $Q$ and $\mu$. Alternatively,
we have also done numerical calculations fixing the charge $Q$ which would then uniquely determine the values of $\psi(0)$, $\psi_2$ and $\mu$. 

Here, we can also redefine the metric function $f(r)$ in a Schwarzschild-like way
\begin{equation}
 f(r)=1-\frac{2m(r)}{r} + \frac{r^2}{\ell^2} \ ,
\end{equation}
where $m(r)$ is the mass function. It is well known that the gravitational mass $M_{\rm G}$ can then be 
expressed easily in terms of this
mass function as
\begin{equation}
 M_{\rm G}=\frac{1}{16\pi G}\lim_{r\rightarrow \infty} m(r) \ .
\end{equation}

\subsection{Minimally coupled scalar field}
This corresponds to the case $\beta=0$.

\subsubsection{Solitons with vanishing charge}
\label{scaling}
Following an argument in \cite{hertog_horowitz} it is easy to show that the gravitational mass $M_G$ of our solutions
with $\phi\equiv 0$
behaves as follows under a rescaling of the form $r\rightarrow \mu r$, with $\mu$ a constant:
\begin{equation}
 M_{G}^{(\mu)} = \frac{M_1}{\mu^3} + \frac{M_2}{\mu}
\end{equation}
where
\begin{equation}
 M_1=\int\limits_0^{\infty} \exp\left(-\frac{1}{2}(\int\limits_{\tilde{r}}^r d\tilde{r} \tilde{r} 
\psi'^2)\right)\left(2\lambda \psi^4 + \frac{6}{\ell^2}\right) \tilde{r}^2 d\tilde{r} \ \ , \ \ 
M_2=\int\limits_0^{\infty} \exp\left(-\frac{1}{2}(\int\limits_{\tilde{r}}^r d\tilde{r} \tilde{r} 
\psi'^2)\right)\left(1+\frac{\tilde{r}^2}{\ell^2}\right)\psi'^2 \tilde{r}^2 d\tilde{r} \ .
\end{equation}
Now $M_2$ is fundamentally positive, however, $M_1$ can become negative for $\lambda$ ``small enough''.
The uncharged soliton must fulfill
\begin{equation}
 \left.\frac{dM}{d\mu}\right\vert_{\mu=1}=0 \ \ \ {\rm i.e.} \ \ \ -3 M_1 = M_2  \ .
\end{equation}
This shows clearly that for $\lambda$ positive uncharged soliton solutions in AdS do not
exist. However, if $\lambda$ is chosen ``negative enough'' $M_1$ can become equal to $-M_2/3$. We observe
the existence of these type of solutions and give more details in the numerical results section below.
The question might than be whether negative values of $\lambda$ are physical since the potential
is not bounded from below. For this we refer the reader to \cite{menagerie}, where different
potentials have been discussed in the context of solitons in AdS. In particular, a potential
of the form $V(\psi)=-2(2+\cosh(\sqrt{2}\psi))= -6 -2\psi^2 - \frac{1}{3}\psi^4 +{\cal O}(\psi^6)$ has been
discussed. 
This potential clearly has a negative coupling in front of the $\psi^4$ term. Here, we want to
study conformally invariant models and since the $\psi^4$ term is the only conformally invariant potential
term, we restrict to this here.

\begin{figure}[h!]
\begin{center}
\subfigure[][]{\label{M_psi0_minimal}\includegraphics[width=13.0cm]{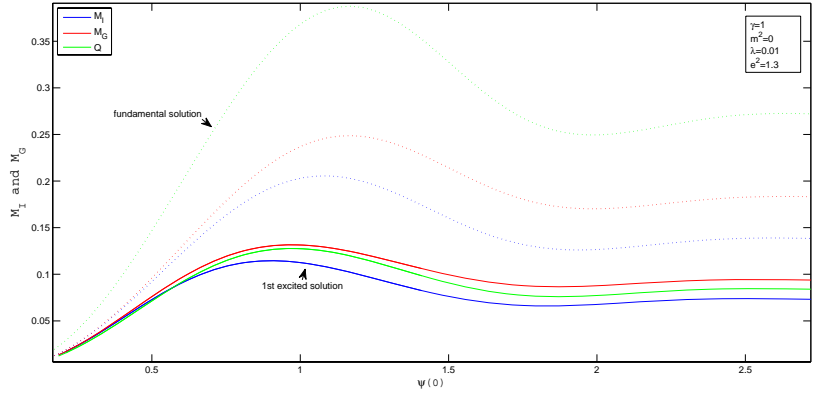}}
\subfigure[][]{\label{M_Q_minimal}\includegraphics[width=13.0cm]{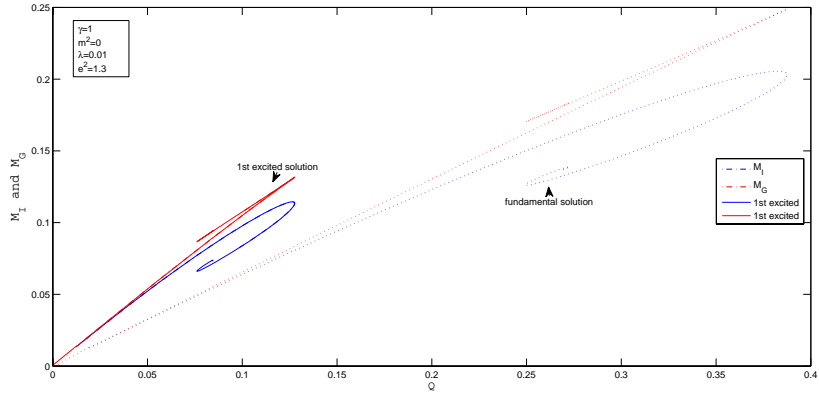}}
\end{center}
\caption{We show the inertial mass $M_I$ (blue), the gravitational mass $M_G$ (red) and the charge $Q$ (green)
as function of $\psi(0)$ for $e^2=1.3$ and $\lambda=0.01$ (a) as well as $M_G$ and $M_I$ as functions
of $Q$ (b). The dotted lines correspond to the fundamental
solutions with no nodes of the scalar field function $\psi(r)$, while we also show the 
results for the first excited solutions with one node of $\psi(r)$ (solid). 
For the dotted lines in (a) the upper curve corresponds to $Q$, while the
lowest curve to $M_I$, respectively. For the solid lines in (a) the upper curve
corresponds to $M_G$, while the lowest corresponds to $M_I$.
  \label{fig_minimal}}
\end{figure}

\begin{figure}[h!]
\begin{center}
\subfigure[][]{\label{M_psi0_minimal2}\includegraphics[width=13.0cm]{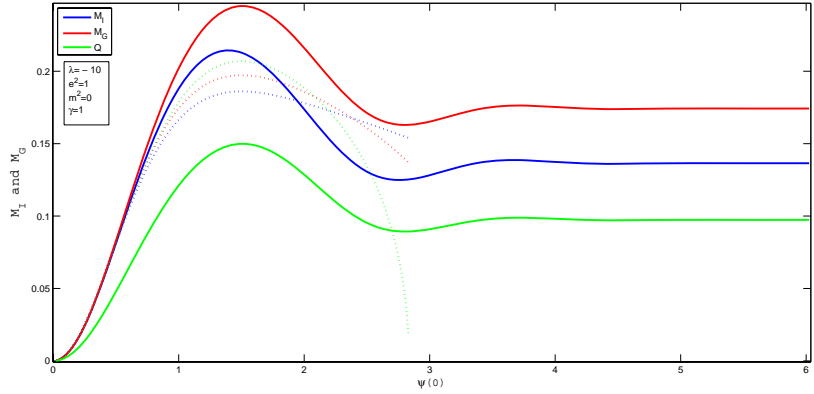}}
\subfigure[][]{\label{M_Q_minimal2}\includegraphics[width=13.0cm]{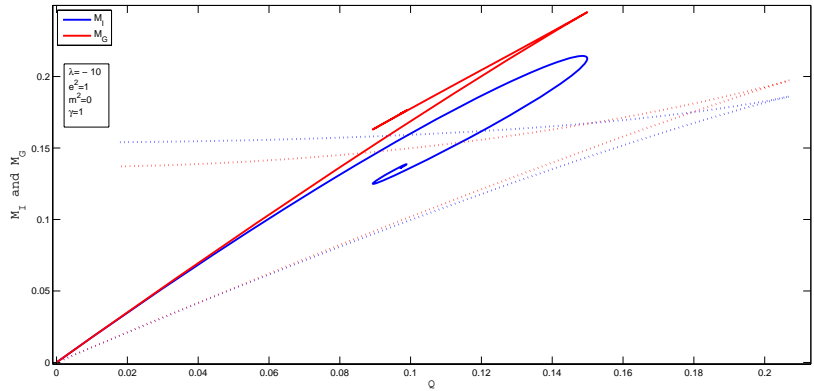}}
\end{center}
\caption{We show the inertial mass $M_I$ (blue), the gravitational mass $M_G$ (red) and the charge $Q$ (green)
as function of $\psi(0)$ for $e^2=1$ and $\lambda=-10$. The dotted lines correspond to the fundamental
solutions with no nodes of the scalar field function $\psi(r)$, while we also show the 
results for the first excited solutions with one node of $\psi(r)$ (solid).For the dotted lines in (a) the upper curve corresponds to $Q$, while the
                               lowest curve to $M_I$, respectively. For the solid lines in (a) the upper curve
                               corresponds to $M_G$, while the lowest corresponds to $Q$. For both dotted and
                               solid linesin (b) the respective curve with the largest value for a given $Q$ corresponds
                               to $M_G$.                               
  \label{fig_minimal2}}
\end{figure}

\begin{figure}[h!]
\begin{center}
\subfigure[][$\lambda=-5.1602 $]{\label{lam_gt_lamcr}\includegraphics[width=11.0cm]{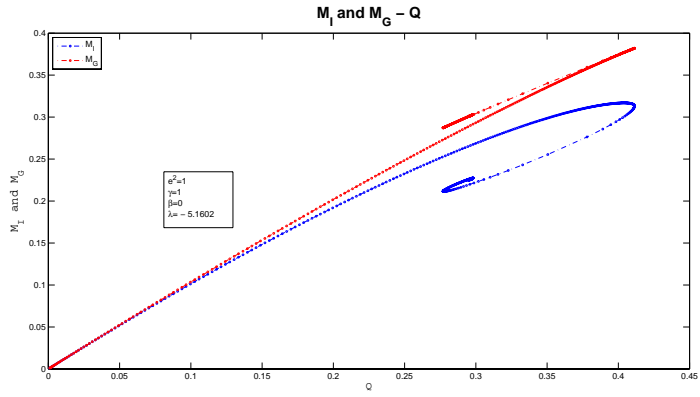}}\\
\subfigure[][$\lambda=-5.4$]{\label{lam_eq_lamcr}\includegraphics[width=11.0cm]{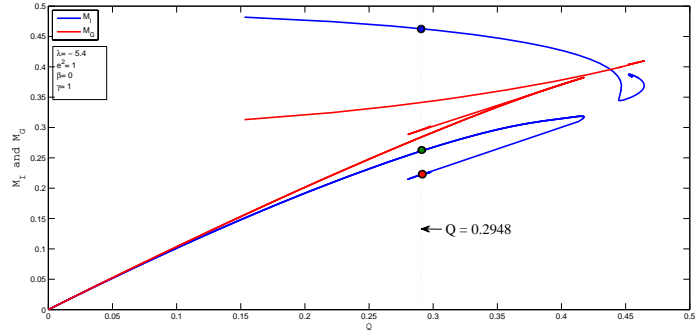}}\\
\subfigure[][$\lambda=-5.44$]{\label{lam_lt_lamcr}\includegraphics[width=11.0cm]{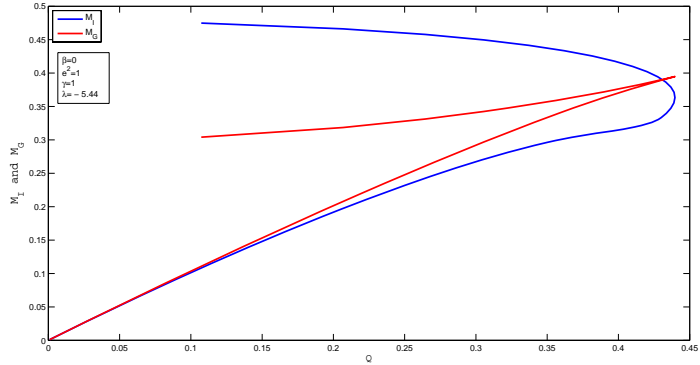}}\\
\end{center}
\caption{We show the gravitational mass $M_G$ (red) and the inertial mass $M_I$ (blue) as function of $Q$ 
for $e^2=1$, $\beta=0$ and different values of $\lambda$.  The three dots in (b) indicate the three possible
solutions for $Q=0.2948$ the profiles of which are given in Fig.\ref{figcomparison}.
In all cases, the curve that possesses a spikelike behaviour corresponds to $M_G$.
  \label{lamcr}}
\end{figure}

\begin{figure}[h!]
\centering
\leavevmode\epsfxsize=14.0cm
\epsfbox{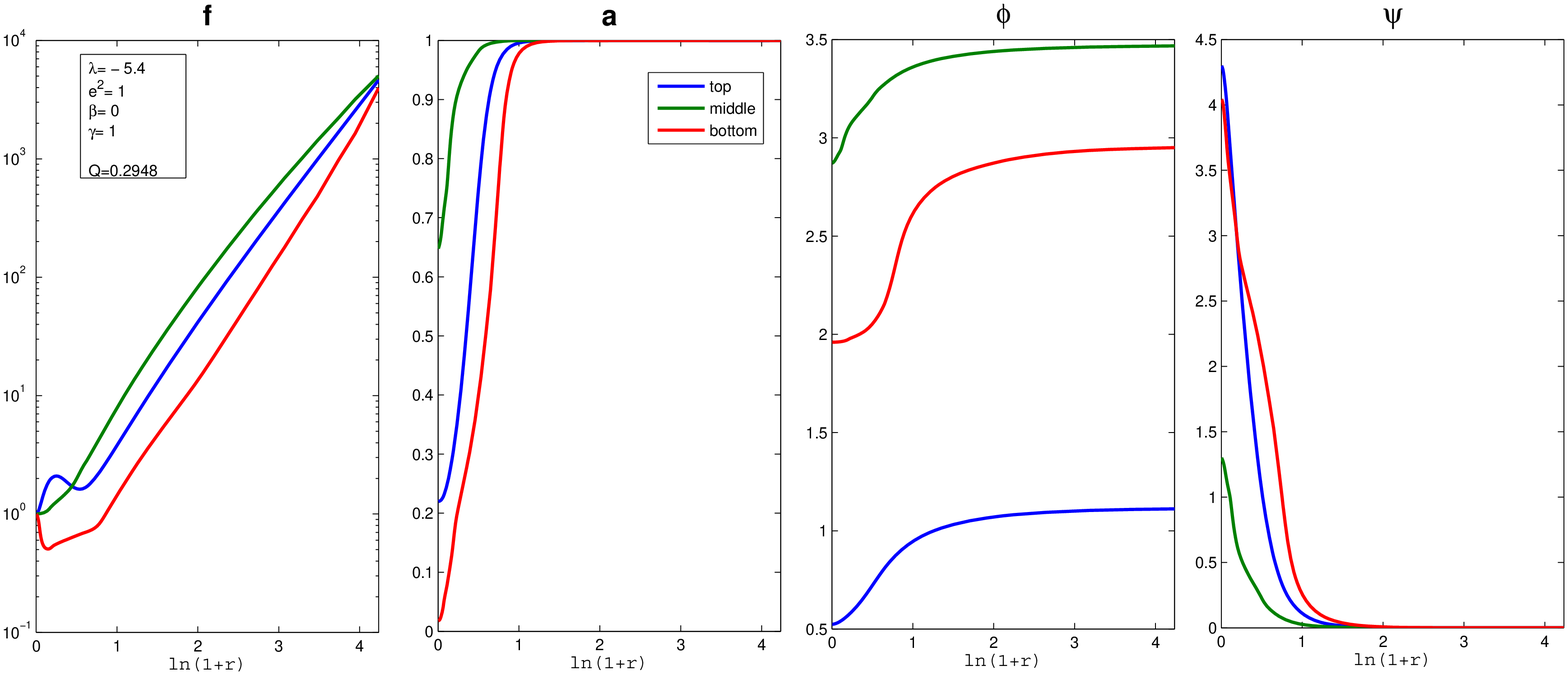}\\
\caption{\label{figcomparison}  
\small{We show the profiles of the solutions existing for a fixed value of $Q=0.2948$ and $e^2=1$, $\beta=0$
and $\lambda=-5.4$.}}
\end{figure}

\subsubsection{Numerical results}

We have used a Newton-Raphson algorithm with adaptive grid scheme \cite{colsys} to construct the solutions numerically.
In the numerical construction, we set $\ell^2=1$ and $\gamma\equiv 16\pi G=1$ without loss of generality. 

In this section, we would like to point out the effect of different choices of $\lambda$. As indicated above $Q=0$ solutions are
not excluded for a choice of $\lambda$ ``negative enough''. This is what we will demonstrate below. First, we have however studied the
case of positive $\lambda$. In Fig.\ref{fig_minimal} we present our results for $e^2=1.3$ and $\lambda=0.01$. Note that we only present the
branch connected to the AdS vacuum here. We observe that the behaviour of the solutions is qualitatively similar to those with a 
``pure'' mass term \cite{menagerie,dias2}. The solutions exist on a finite interval of the mass and charge. 
We find that both the gravitational as well as the inertial mass have their maximal possible value at the largest possible charge $Q$.
From this value of $Q$ on, a second branch of solutions extends backwards in $Q$. The curve of the gravitational mass $M_G$ simply extends backwards with $M_G$ being higher
on the second branch than on the first branch. The inertial mass on the other hand is lower on the second branch of solutions and shows a spiraling behaviour.
This qualitative behaviour is true for both the fundamental solution
without nodes in the scalar field function $\psi(r)$ as well as for the first excited solution which contains a zero of $\psi(r)$ at some intermediate
value of the coordinate $r$. We also find that the maximal value of the mass and charge is much smaller for the excited solutions than for the fundamental
solutions.

We have then studied the case of negative $\lambda$. Our results for $e^2=1$ and $\lambda=-10$ are shown in Fig.\ref{fig_minimal2}. We observe that very similar
to the case with positive $\lambda$ the AdS vacuum gets smoothly deformed when increasing $\psi(0)$ from zero. The gravitational and inertial mass
as well as the charge $Q$ increase up to a maximal value at some intermediate value of $\psi(0)$. Then all three values decrease such that at some
finite value of $\psi(0)$ the charge of the solution vanishes, i.e. $Q=0$. We find that for our choice of parameters the fundamental solutions hence interpolate between the AdS vacuum and
an uncharged configuration. As can be seen from Fig.\ref{M_Q_minimal2} the qualitative behaviour of the inertial mass in dependence on $Q$
is different for this case and shows a behaviour very similar to that of the gravitational mass. Let us also remark that the first excited solutions
do not seem to tend to a $Q=0$ solution (at least not within the range of $\psi(0)$ that we have studied) and show a qualitative behaviour similar to those of AdS solitons
with positive $\lambda$.

The question is then at which value of $\lambda$ the transition from the qualitative behaviour observed in the $\lambda > 0$ case
to the behaviour observed for $\lambda=-10$ takes place. We would expect this to happen at $\lambda$ ``small enough''. We have hence studied the range of negative
$\lambda$ in more detail. Our results for $e^2=1$ are shown in Fig.\ref{lamcr}. While for $\lambda=-5.1602$ we still find the qualitative behaviour present in the positive
$\lambda$ case, this changes when decreasing $\lambda$ further. For $\lambda=-5.44$ we find a behaviour similar to that for $\lambda=-10$. Our results
for an intermediate value, i.e. $\lambda=-5.4$ clearly show how the transition between the two takes place. We find that the branch showing the inertial mass
of the solutions splits into disconnected ones, i.e. into two branches that are connected to the AdS vacuum and into two branches that connected to the $Q=0$ solution.
The two branches connected to the AdS vacuum look similar to the branches existing for larger values of $\lambda$, while the two branches connected to $Q=0$ seem to
appear as soon as $\lambda$ is small enough. 
Decreasing $\lambda$ we would expect the two branches connected to the AdS vacuum to join the two connected to the $Q=0$ solution to form two branches that interpolate
smoothly between the AdS vacuum and the $Q=0$ solution.

Interestingly, in the case of this critical $\lambda$ where disconnected branches appear, we can have up to three solutions for the same value of $Q$. These three
solutions for $Q=0.2948$ and $\lambda=-5.4$ are shown in Fig.\ref{figcomparison}. The solutions look qualitatively similar, however, we see a clear difference
in the profiles, e.g. the larger the value of $M_I$ the smaller is the value of $\phi(0)$.

\subsection{Conformally coupled scalar field}
This corresponds to $\beta=1/6$. Again, we have used the same numerical technique as before and have fixed 
$\ell^2=1$ and $\gamma\equiv 16\pi G=1$ without loosing generality.

\subsubsection{Numerical results}
We have first studied the gravitational mass $M_G$ and the charge $Q$ of the solutions for different values of $\lambda$ and $e^2=1$.
The results are given in Fig.\ref{all_lambda}. We observe that for positive $\lambda$ the mass
tends to a constant for $\psi(0)\rightarrow \infty$ with solutions with positive $Q$ existing all the way down to 
$a(0)=0$ where they merge with a singular solution. On the other hand for negative $\lambda$ we observe that
for $\lambda$ sufficiently negative the solutions tend to a solution with $Q=0$ at some finite value of
$\psi(0)$ and $a(0)$. These solutions have $\phi\equiv 0$ which can be see in Fig.\ref{approach} for $\lambda=-3$, where
we plot the profiles of the solutions.

\begin{figure}[h!]
\centering
\leavevmode\epsfxsize=14.0cm
\epsfbox{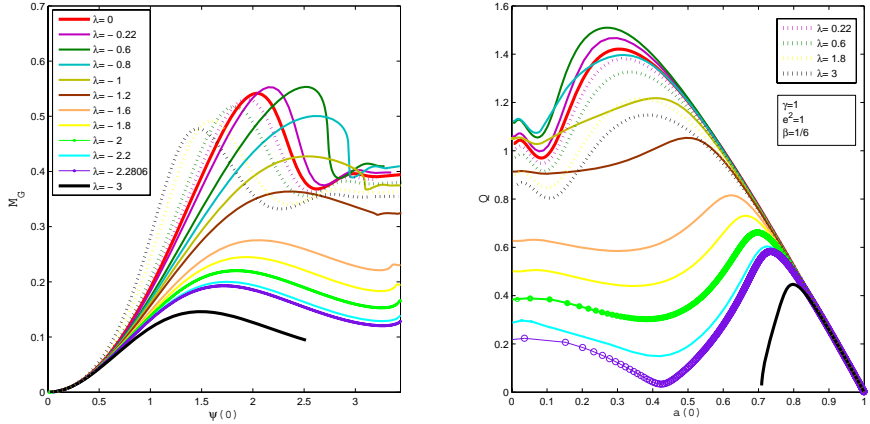}\\
\caption{\label{all_lambda}  
\small{We show the gravitational mass $M_G$ as function of $\psi(0)$ (left) and the charge $Q$ (right) as function of $a(0)$ for $e^2=1$
and different values of $\lambda$ for a conformally coupled scalar field, i.e. for $\beta=1/6$. }}
\end{figure}

\begin{figure}[h!]
\begin{center}
\subfigure[][$M$ and $Q$]{\label{lambda_3}\includegraphics[width=14.0cm]{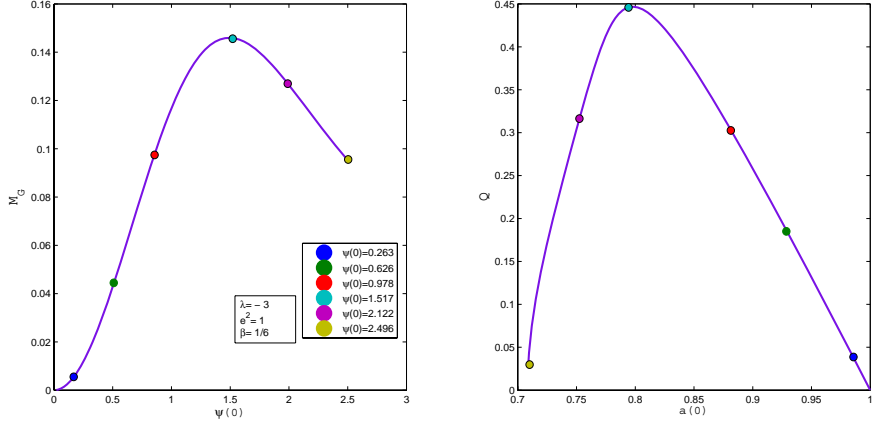}}\\
\subfigure[][profiles]{\label{lambda_3_profiles}\includegraphics[width=14.0cm]{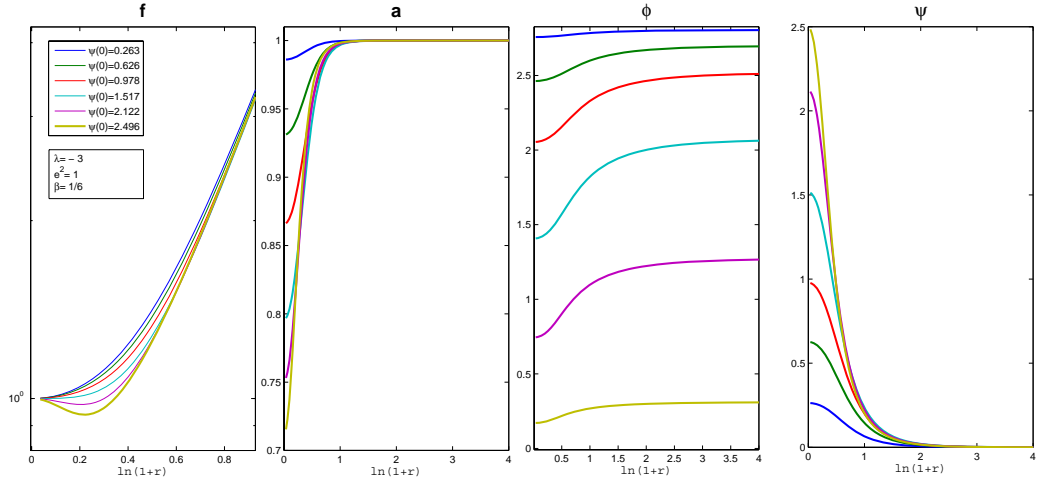}}
\end{center}
\caption{{\it Top}: We show the gravitational mass $M_G$ as function of $\psi(0)$ (left) and the charge $Q$ (right) 
as function of $a(0)$ for $e^2=1$ and $\lambda=-3$. The dots on the curves indicate the solutions
plotted below. {\it Bottom}: We show the profiles of the metric functions $f(r)$, $a(r)$ and of the matter
functions $\phi(r)$ and $\psi(r)$ for $e^2=1$, $\lambda=-3$ and the values of $\psi(0)$ and $a(0)$ as indicated
by the dots in the figure on the top.  \label{approach}}
\end{figure}

\begin{figure}[h!]
\centering
\leavevmode\epsfxsize=14.0cm
\epsfbox{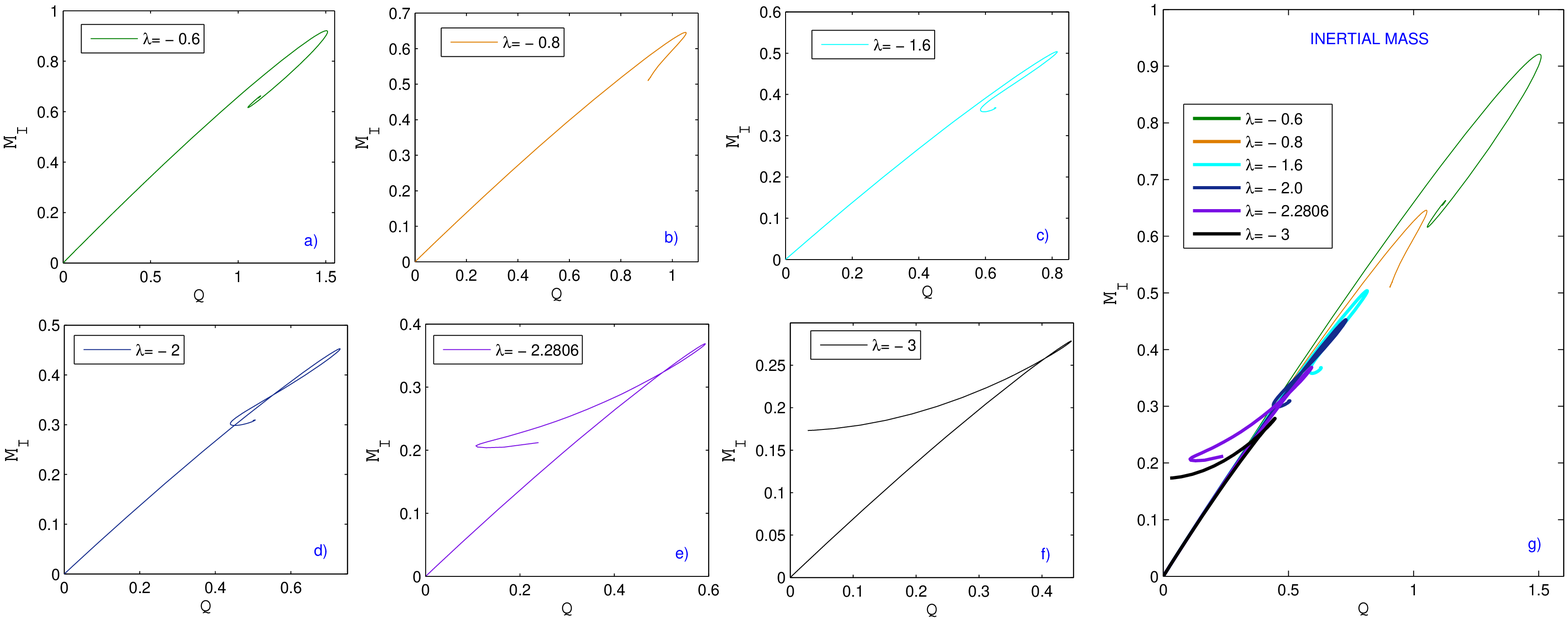}\\
\caption{\label{inertial_mass}  
\small{We show the inertial mass $M_I$  as function of the 
charge $Q$ for $e^2=1$ and different (negative) values of $\lambda$ for a conformally coupled scalar field.}  }
\end{figure}

Since we have observed that the qualitative dependence of the inertial mass on the charge $Q$ changes when
lowering $\lambda$ for $\beta=0$, we have also studied this here. Our results for the dependence of the inertial mass $M_I$ on the charge $Q$ for
$e^2=1$ and different values of $\lambda < 0$ are shown in Fig.\ref{inertial_mass}. We observe that similar to $\beta=0$ the qualitative
features change when choosing $\lambda$ ``small enough''. We observe that for a given $e^2$ this happens at larger $\lambda$ in the case
of a conformally coupled scalar field in comparison to a minimally coupled one. 
This is indicated by Fig. \ref{critical_Q0}, where we give the gravitational mass $M_G$ as function of $\psi(0)$ and the
charge $Q$ as function of $a(0)$ for $e^2=3$ and three different negative values of $\lambda$.
We find that $\lambda\approx -2.2854$ is the largest value at which $Q=0$ solutions are possible. There is just one
solution with $Q=0$ in this case. Lowering $\lambda$ further, we find that two solutions with $Q=0$ can exist
which possess different values of $a(0)$, $\psi(0)$ and the gravitational mass $M_G$. Furthermore, we find
that a ``gap'' opens up in between these two $Q=0$ solutions.

\begin{figure}[h!]
\centering
\leavevmode\epsfxsize=14.0cm
\epsfbox{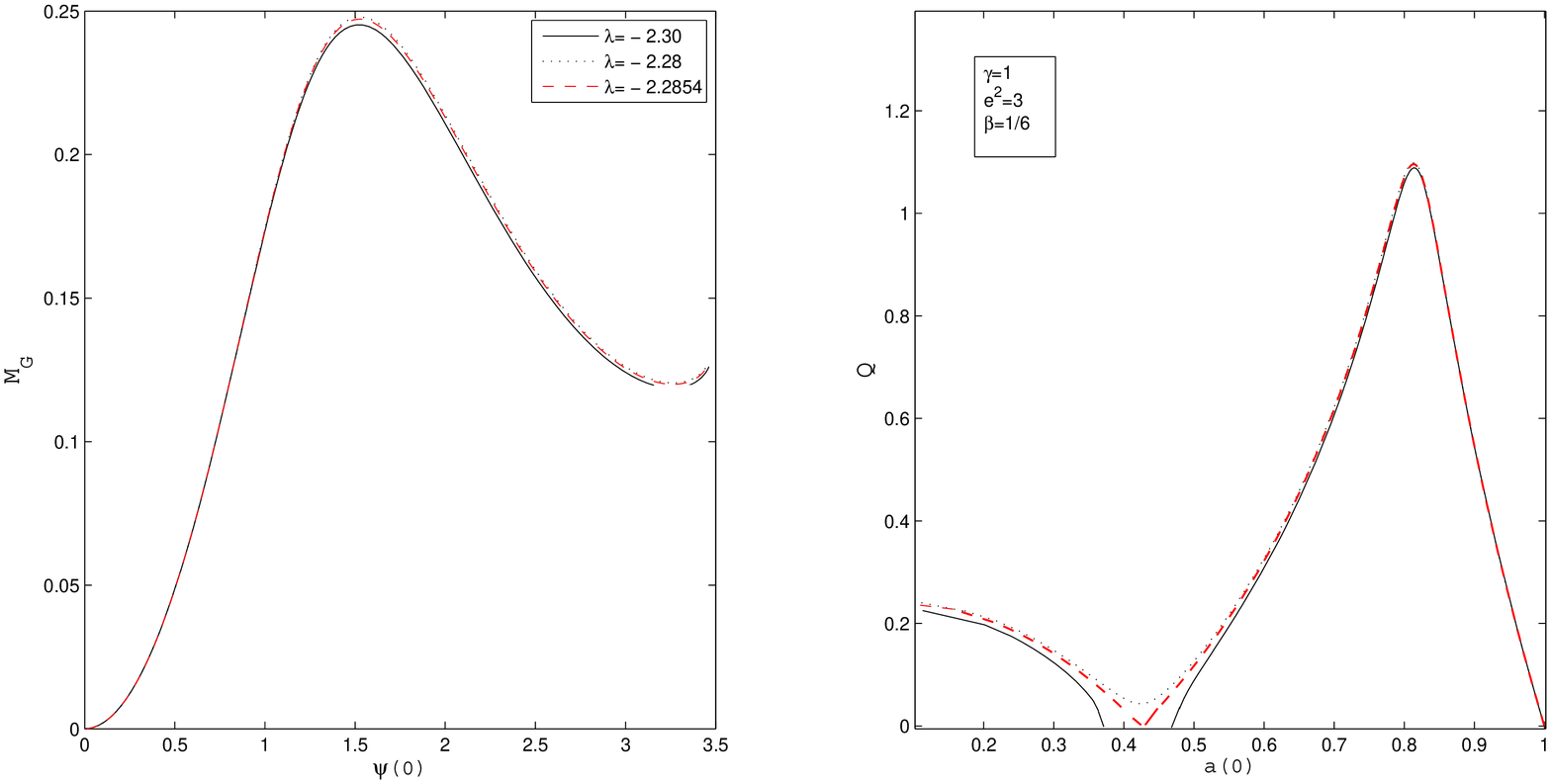}\\
\caption{\label{critical_Q0}  
\small{We show the gravitational mass $M_G$  as function of $\psi(0)$ (left) as well as the charge $Q$ 
as function of $a(0)$ (right) for $e^2=3$, $\beta=1/6$ and different values of negative
$\lambda$. }}
\end{figure}

\section{Conformal gravity}
In the following, we want to study a model that is conformally invariant both in the matter as well as in the gravity part.
As stated above, the Einstein-Hilbert action is not conformally invariant. We hence replace this by a conformally invariant action
whose main ingredient is the Weyl tensor $C_{\kappa\lambda\mu\nu}$ given by the totally traceless part of the Riemann tensor
\begin{eqnarray}
C_{\kappa\lambda\mu\nu}=R_{\kappa\lambda\mu\nu}-
\frac{1}{2}(g_{\kappa\mu}R_{\lambda\nu}-g_{\kappa\nu}R_{\lambda\mu}+
g_{\lambda\nu}R_{\kappa\mu}-g_{\lambda\mu}R_{\kappa\nu})+
\frac{R}{6}(g_{\kappa\mu}g_{\lambda\nu}-g_{\kappa\nu}g_{\lambda\mu})
\label{WeylTensor}   \ .
\end{eqnarray}
The gravitational Lagrangian then reads
\begin{equation}
{\cal L}_{\rm gravity}= -\frac{1}{2\alpha}C_{\kappa\lambda\mu\nu}C^{\kappa\lambda\mu\nu} 
\label{GravL}
\end{equation}
where $\alpha$ is a dimensionless positive parameter. The matter field equations are given
again by (\ref{FieldEqsScalar}) and (\ref{FieldEqsVector}). The gravitational field equations are 
formally similar to the Einstein equations and are given by
\begin{equation}
W_{\mu\nu} =  \frac{\alpha}{2} T_{\mu\nu} 
\label{GravFieldEq}  \ ,
\end{equation}
where $W_{\mu\nu}$ is the Bach tensor defined by:
\begin{eqnarray}
W_{\mu\nu}=\frac{1}{3}\nabla_\mu\nabla_\nu R-\nabla_\lambda\nabla^\lambda R_{\mu\nu}
+\frac{1}{6} (R^2+\nabla_\lambda\nabla^\lambda R-3R_{\kappa\lambda}R^{\kappa\lambda})g_{\mu\nu}+
2R^{\kappa\lambda}R_{\mu\kappa\nu\lambda}-\frac{2}{3}RR_{\mu\nu}
\label{BachTensor}
\end{eqnarray}
Since the Bach tensor is traceless, the energy-momentum tensor $T_{\mu}^{\nu}$  must also fulfill $T^\mu_\mu=0$.
It is important to note that the gravity Lagrangian does not contain a negative cosmological constant a priori. As we
will see later, a negative cosmological constant is naturally induced in conformal gravity due to appropriate
choices of the boundary conditions. 

We again use the Ansatz (\ref{ansatz_metric}) and (\ref{ansatz_matter}), however the conformal symmetry allows us
to set $a(r)\equiv 1$ \cite{Mannheim2006}. 

The field equations for the matter fields are given by (\ref{eq3}) and (\ref{eq4}), while
the gravity equation reads
\begin{equation}
\frac{f(rf)''''}{r} = - \frac{3 \alpha}{2} (T_0^0 - T_r^r)
\label{gravity_equation}
\end{equation}
with ${T}_{\mu\nu}$ the energy-momentum tensor that results from the 
variation of the matter Lagrangian with respect to the metric:
\begin{equation}
{T}_{\mu\nu} = { \tilde T}_{\mu\nu}+{1\over 3}
\left (g_{\mu \nu} \nabla ^{\lambda }\nabla_
{\lambda } |\psi|^2 - \nabla _{\mu }\nabla_{\nu } |\psi|^2  - 
{G}_{\mu \nu} |\psi|^2 \right)
\label{confTmn}
\end{equation}
where ${\tilde T}_{\mu\nu}$ represents the ordinary (``minimal'') energy-momentum tensor of the 
U(1) scalar field model and ${G}_{\mu \nu}$ is the Einstein tensor. The non-vanishing components of this
tensor are given e.g. in \cite{BrihayeVerbin}.

To solve the system of differential equation that are now fourth order in the gravitational sector 
we choose the following boundary conditions 
\be
f(0)=1 \ , \ f'(0) = 0 \ , \ f'''(0)=0 \ , \ \phi'(0) = 0 \ , \ \psi'(0) = 0 \ , 
\ee
\be 
 f''(r\to \infty) = 2 \kappa
\ , \ (r^2 \phi'(r))_{r \to \infty} = Q \ , \ (r^2 \psi'(r))_{r \to \infty} = 0  \ ,
\ee  
where $\kappa$ and $Q$ are constants, where the former corresponds to the effective negative
cosmological constant and the latter is the charge.

The system of coupled differential equations has similar scaling symmetries than the one in the 
Einstein gravity case. These read
\begin{equation}
 r\rightarrow \sigma r \ \ , \ \  t\rightarrow \sigma t \ \ , \ \ \kappa \rightarrow \sigma \kappa \ \ , \ \ e\rightarrow e/\sigma \ \ , \ \
\lambda \rightarrow \lambda/\sigma^2 \ ,
\end{equation}
\begin{equation}
 \psi\rightarrow \sigma \psi \ \ , \ \  \phi\rightarrow \sigma \phi \ \ , \ \ \alpha \rightarrow \alpha/\sigma^2 \ \ , \ \ e\rightarrow e/\sigma \ \ , \ \
\lambda \rightarrow \lambda/\sigma^2 \ 
\end{equation}
and can be used to set two coupling constants to fixed values without loosing generality. 

Here, it was shown that the metric function $f(r)$ has the general behaviour far away from sources \cite{BrihayeVerbin}
\begin{equation}
\label{f_infty}
 f(r>>1)=\kappa r^2 + c_0 + c_1 r + \frac{c_2}{r} \ ,
\end{equation}
where the $c_i$, $i=0,1,2$ are integration constants and $\kappa$ is the aforementioned cosmological constant. Using 
(\ref{gravity_equation}) it is then apparent that
\begin{equation}
 M_G= \frac{16\pi c_1}{\alpha} \ .
\end{equation}
Hence, the gravitational mass is related to the linear term in (\ref{f_infty}). This is not surprising since
the potential of a point particle in conformal gravity is linear. The coefficient $c_2$ on the other hand,
which determines the $1/r$-fall off of the metric function and in Einstein-Hilbert gravity would
be interpreted as the mass can be expressed as follows \cite{BrihayeVerbin}
\begin{equation}
 c_2= \frac{\alpha}{4} \int\limits_0^{\infty}  dr \ r^4 \ \left(T^0_0(r) - T^r_r(r)\right)/f(r)  \ .
\end{equation}
In the following, we will determine $c_2$ as well as $M_{\rm G}$.

\subsection{Numerical results}
We solved the system of field equations numerically using the same technique as above \cite{colsys}. In the following we will choose $\alpha=1$ and $\kappa = 0.1$.
The relevant  parameters to be considered are therefore $e^2$ and $\lambda$.

\begin{figure}[h!]
\begin{center}
\includegraphics[width=13.0cm]{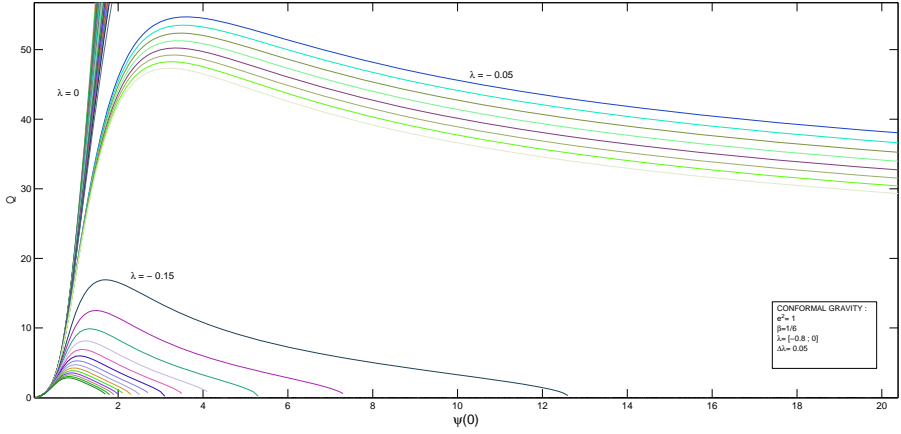}
\end{center}
\caption{We show the charge $Q$  as function
of $\psi(0)$ for $e^2=1$ and different values 
of $\lambda \le 0$ in the case of conformal gravity and $\beta=1/6$. Note that the qualitative features
for $\lambda > 0$ are similar to those for $\lambda=0$.
  \label{M_Q_psi0_conformal}}
\end{figure}

\begin{figure}[h!]
\begin{center}
\subfigure[][]{\label{M_psi_lam_conf}\includegraphics[width=13.0cm]{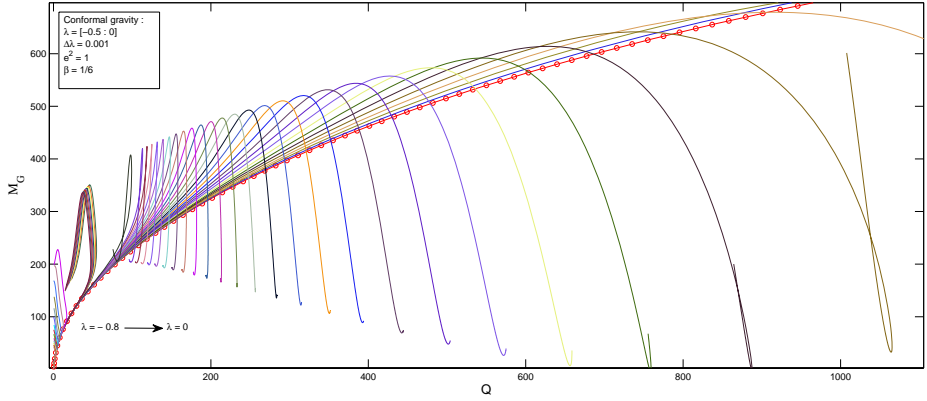}}
\subfigure[][]{\label{M_Q_lam_conf}\includegraphics[width=13.0cm]{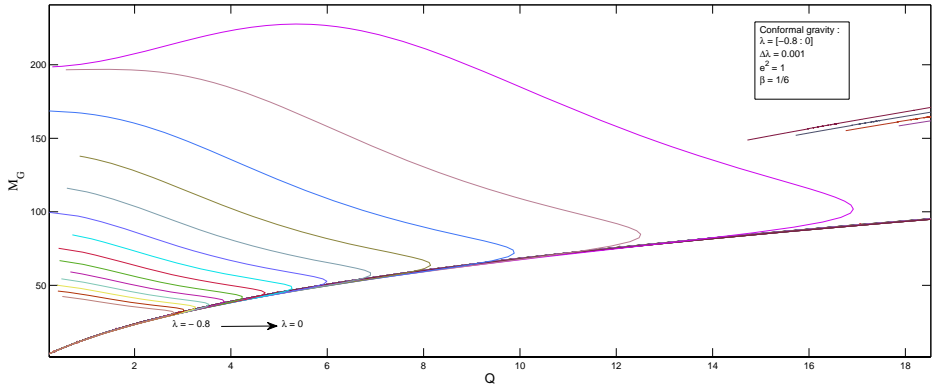}}
\end{center}
\caption{We show the gravitational mass $M_G$ as function of $Q$ for $e^2=1$ and different values 
of $\lambda\in [-0.8:0]$ in the case of conformal gravity and $\beta=1/6$. Going from left to right
the value of $\lambda$ associated to the curves increases in steps of $\Delta\lambda=0.001$.
The upper figure (a) shows the whole range of data that we obtained, while the lower figure (b) 
shows a zoom onto the region with $Q\in [0:18]$. 
  \label{lambda_conf}}
\end{figure}

We have first studied the dependence of the qualitative pattern on $\lambda$ for a fixed value of $e^2$. 
Our results for
$e^2=1$ are shown in Fig.\ref{M_Q_psi0_conformal}. 
We find that for a given value of $\psi(0)$ the charge $Q$ 
decrease when decreasing $\lambda$ from zero \footnote{Note that the qualitative features for $\lambda > 0$ 
are similar to those in the $\lambda=0$ case. This is why we do not 
discuss them in detail here.}. We find that for $\lambda$ close to zero the charge increases strongly with $\psi(0)$ suggesting
that the solutions reach the planar limit at large enough $\psi(0)$. The behaviour changes when choosing $\lambda$ 
smaller,
in our case for $\lambda \le -0.05$. The charge $Q$ now reaches a constant, finite value for $\psi(0)\rightarrow \infty$.
For even smaller values of $\lambda$ we observe the same phenomenon as for the Einstein gravity case, namely
that the solutions tend to a $Q=0$ solution at some finite value of $\psi(0)$. This is clearly seen in Fig.\ref{M_Q_psi0_conformal}
for $\lambda \le -0.15$.

Furthermore, we observe that the maximal value of 
the gravitational mass is
reached at smaller and smaller values of $Q$ when decreasing $\lambda$, see Fig.\ref{lambda_conf}.
We also do not observe a 
spike-like behaviour in the $M_G$-$Q$-plot which means
that the maximal gravitational mass in not reached at the maximal charge $Q$. Moreover, the 
qualitative pattern seems to change when varying $\lambda$. For $\lambda$ close to zero the gravitational
mass $M_{\rm G}$ has a maximum, then decreases strongly on a second branch of solutions that have larger $Q$, forms a loop and increases again. When decreasing
$\lambda$ further, the loop shrinks in size and the minimal value of $M_{\rm G}$ decreases. Our numerical
results indicate that $M_{\rm G}$ can even become negative for some values of $Q$. For even smaller values of $\lambda$
we observe that the loop disappears, that the minimal value of $M_{\rm G}$ increases again until a second
branch of solutions forms that extends back to smaller values of $Q$. As can be clearly seen in the zoom
on the domain $Q\in[0:18]$ in Fig.\ref{lambda_conf} this second branch extends all the way back to $Q=0$ 
if $\lambda$ is small enough. All our numerical results indicate that a solution with $Q=0$ also exists in the
case of conformal gravity. However, applying a scaling-type argument similar to that in (\ref{scaling}) does not seem
to work here. If we would apply this to either $M_{\rm I}$, $M_{\rm G}$ or $c_2$ we would always
find that these quantities should be zero. Hence, other arguments have to be evoked in conformal gravity
in order to show the existence of soliton-type solutions.

\begin{figure}[h!]
\begin{center}
\subfigure[][]{\label{M_I_Q_small_e}\includegraphics[width=13.0cm]{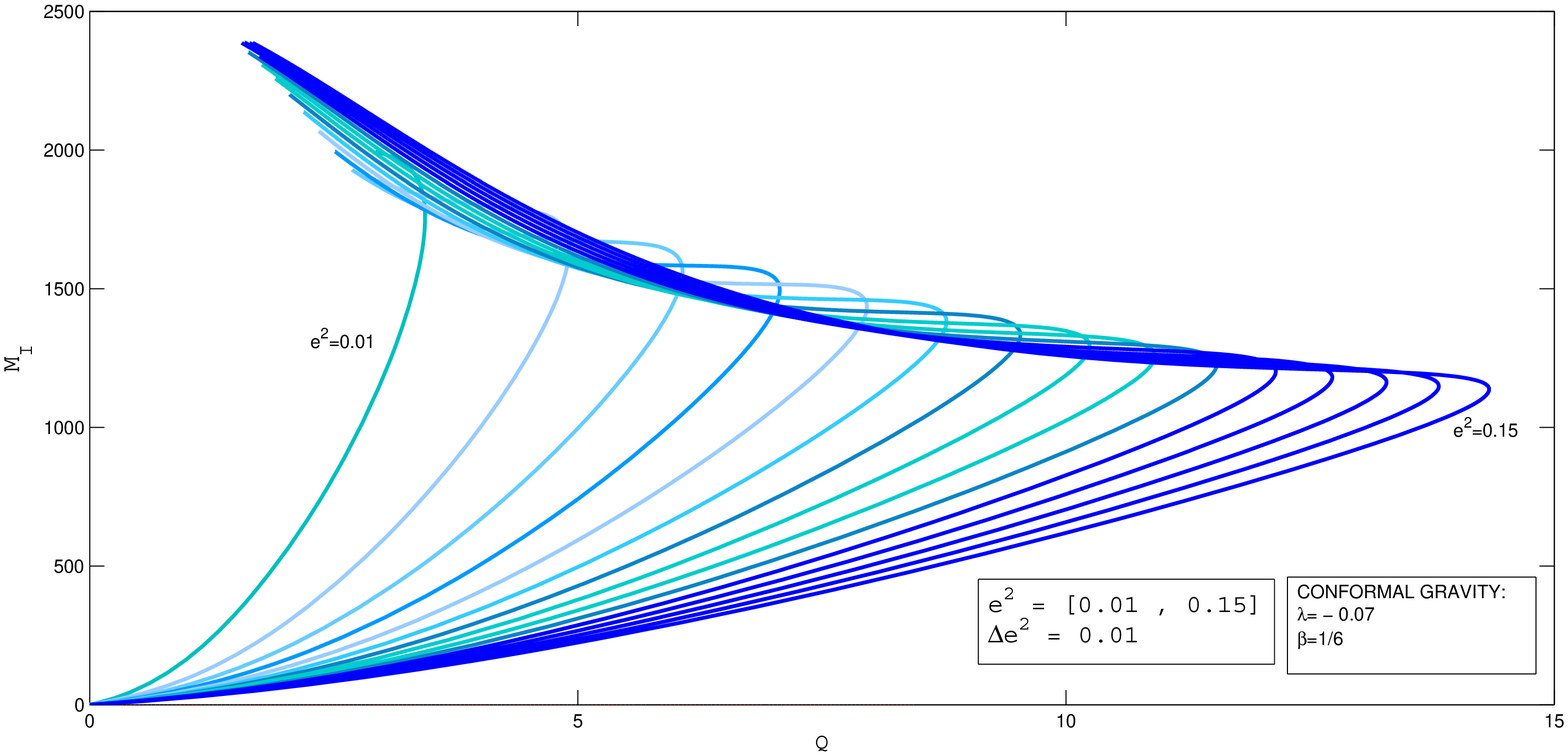}}
\subfigure[][]{\label{M_G_Q_small_e}\includegraphics[width=13.0cm]{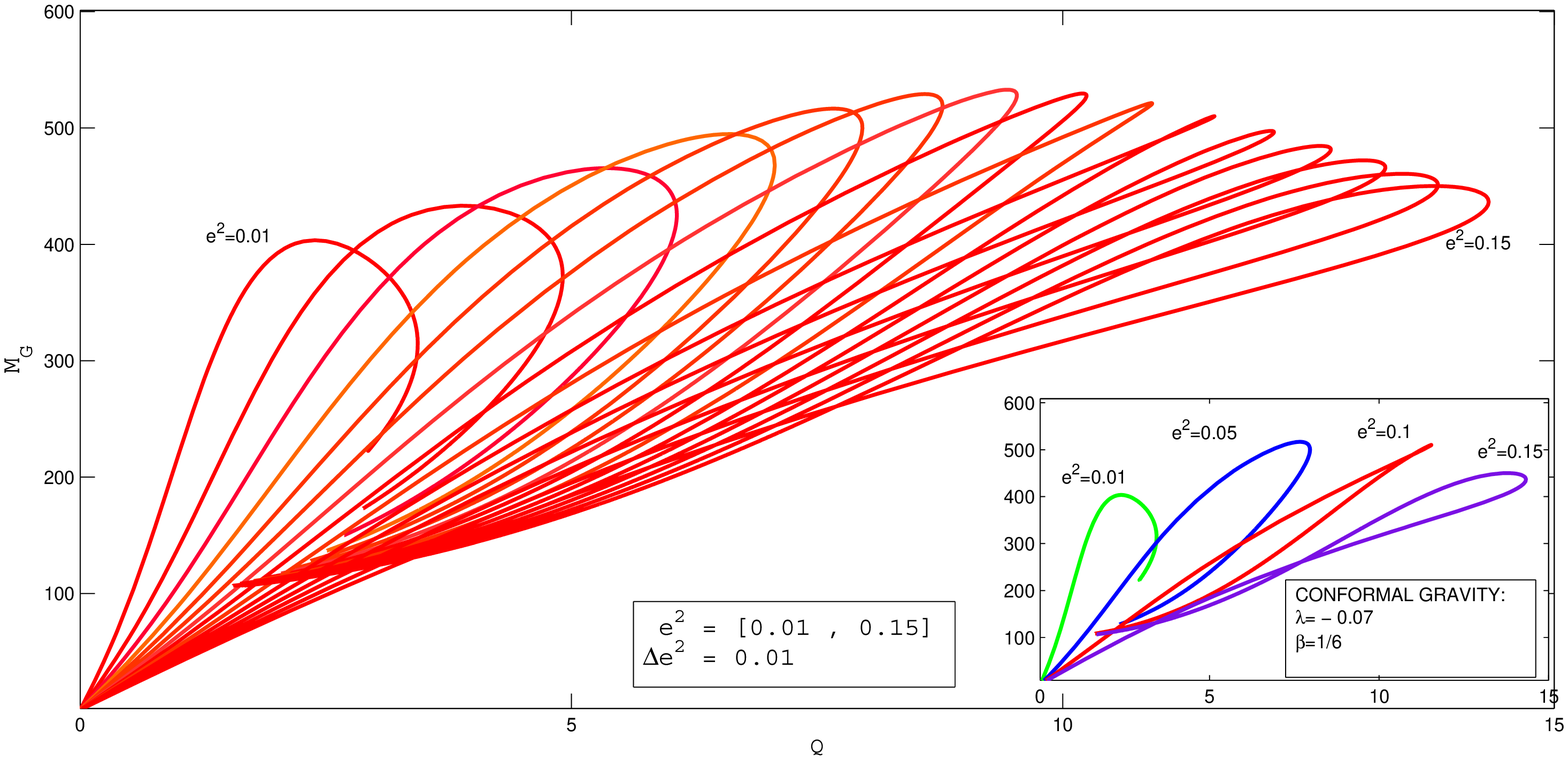}}
\subfigure[][]{\label{c_2_Q_small_e}\includegraphics[width=13.0cm]{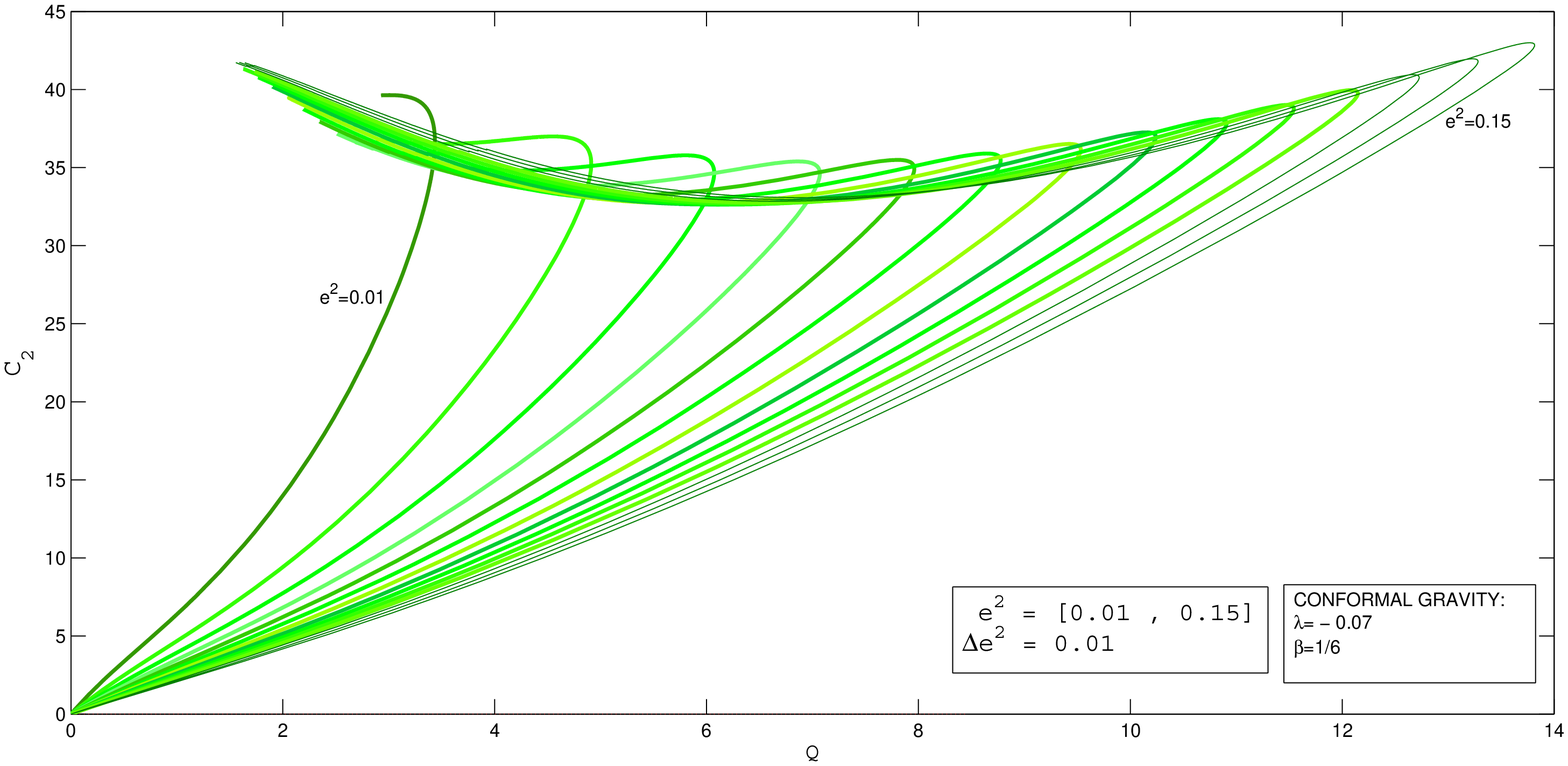}}
\end{center}
\caption{We show the inertial mass $M_I$ (a), the gravitational mass $M_G$ (b) and the parameter $c_2$ (c), respectively as function of
$Q$ for different values of $e^2\in [0.01:0.15]$ and $\lambda=-0.07$
in the case of conformal gravity and $\beta=1/6$.
  \label{small_e}}
\end{figure}

\begin{figure}[h!]
\begin{center}
\subfigure[][]{\label{M_I_Q_large_e}\includegraphics[width=13.0cm]{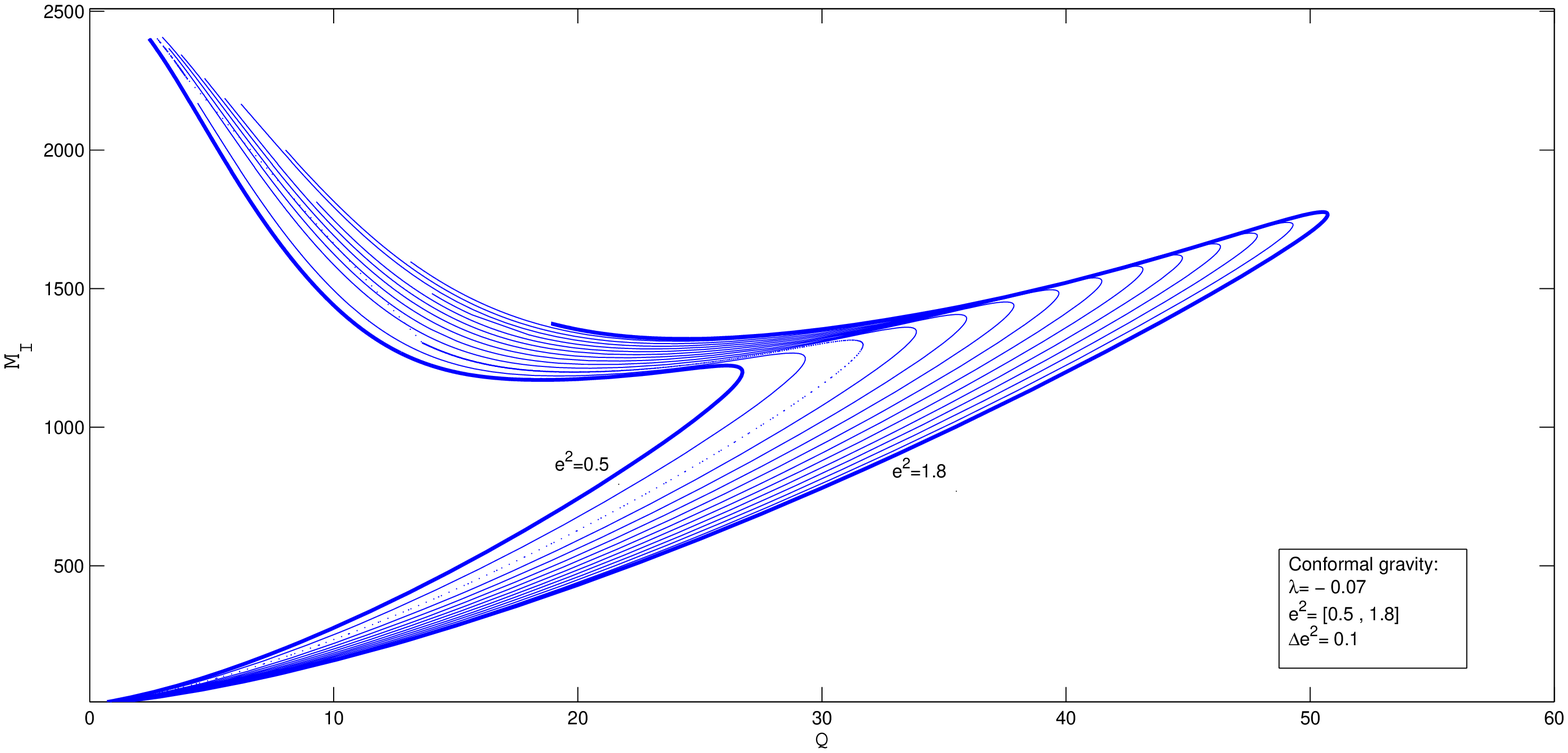}}
\subfigure[][]{\label{M_G_Q_large_e}\includegraphics[width=13.0cm]{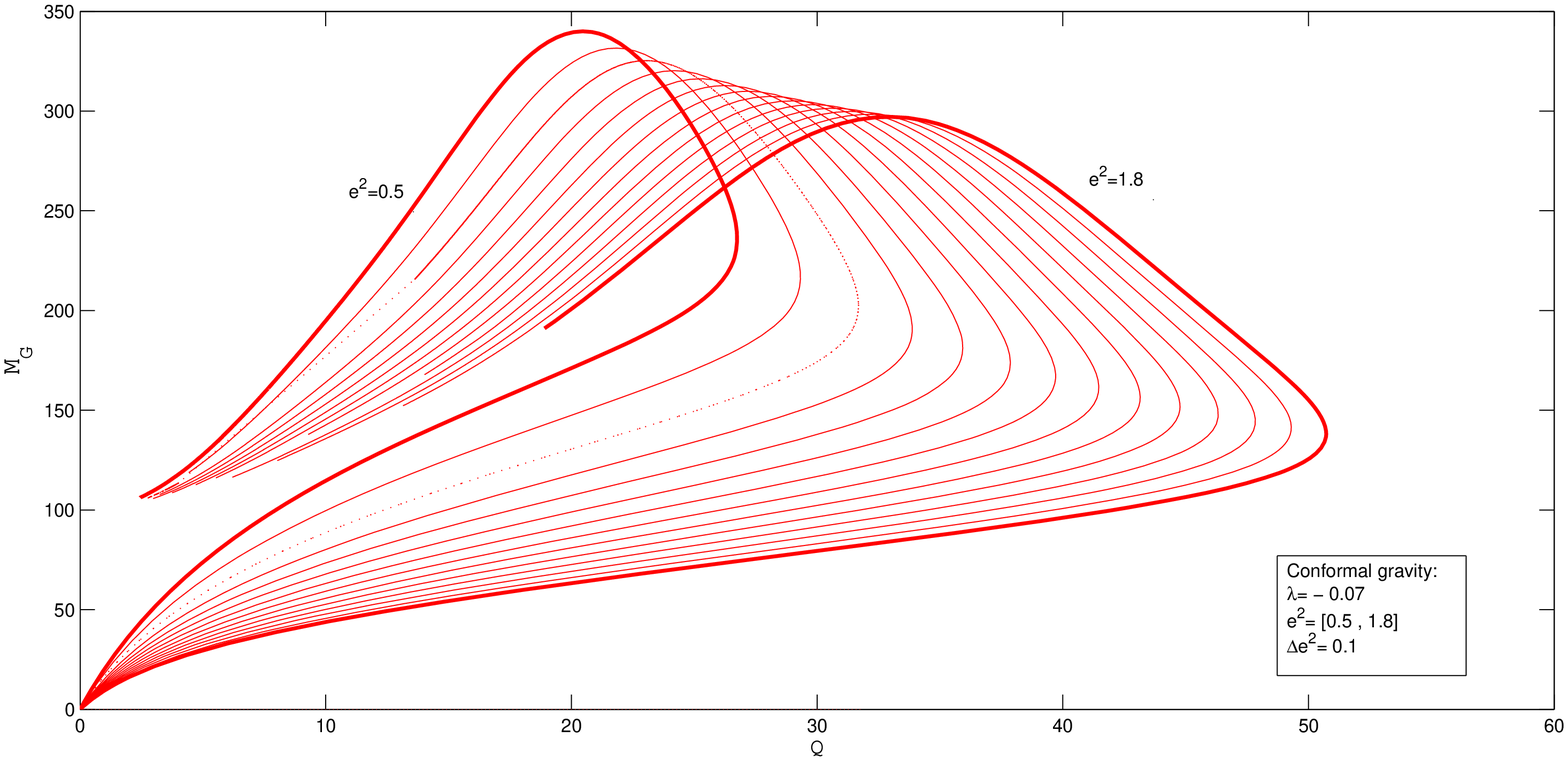}}
\subfigure[][]{\label{c_2_Q_large_e}\includegraphics[width=13.0cm]{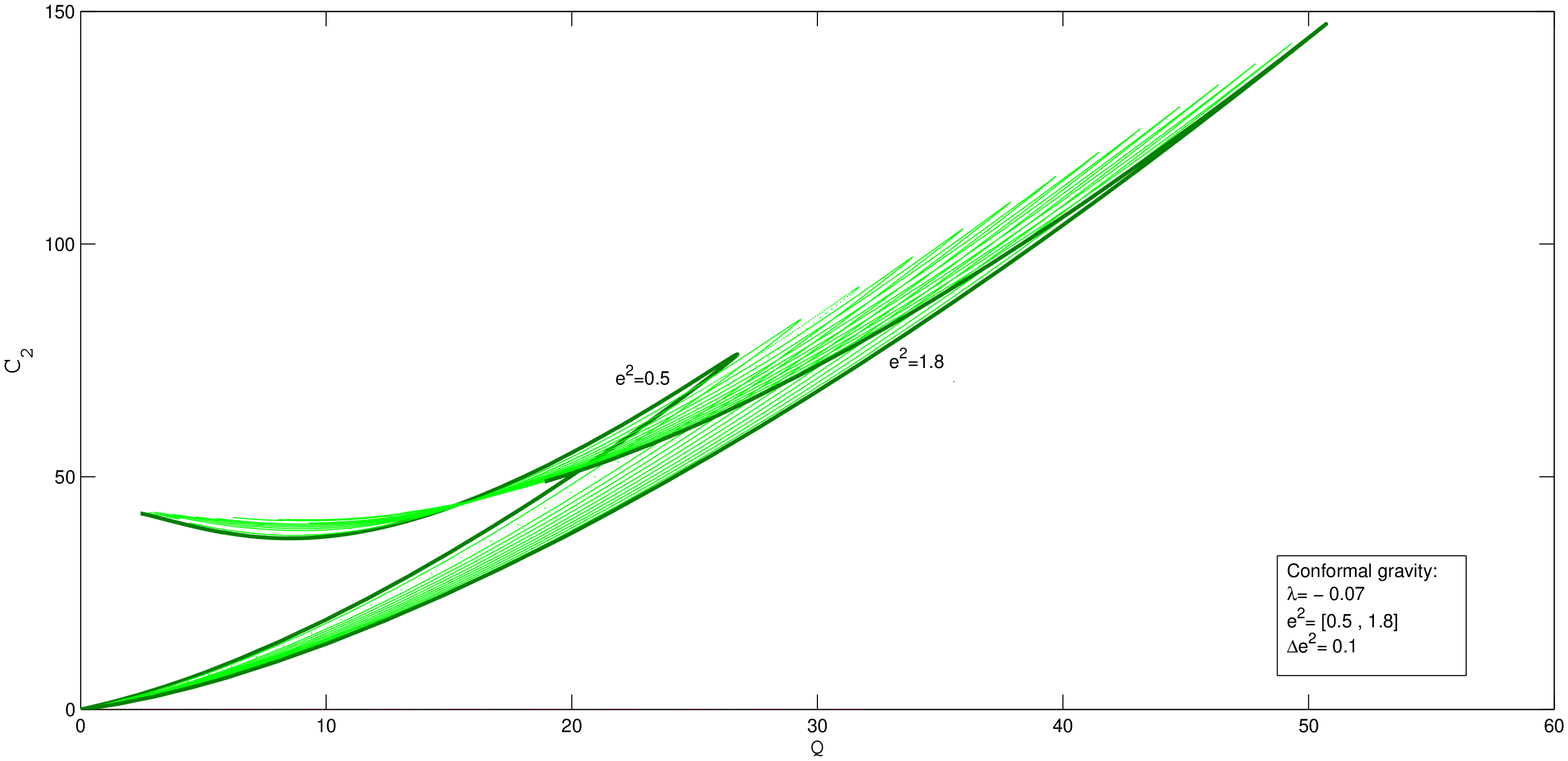}}
\end{center}
\caption{We show the inertial mass $M_I$ (a), the gravitational mass $M_G$ (b) and the parameter $c_2$ (c), respectively as function of
$Q$ for different values of $e^2\in [0.5:1.8]$ and $\lambda=-0.07$
in the case of conformal gravity and $\beta=1/6$.
  \label{large_e}}
\end{figure}

In Fig. \ref{small_e} and Fig.\ref{large_e} we give the inertial mass $M_{\rm I}$, 
the gravitational mass $M_{\rm G}$
and the parameter $c_2$ as function of $Q$ for $\lambda=-0.07$ and varying values of $e^2$.
The first thing to note is that the qualitative pattern seems to be similar when comparing the inertial
mass $M_{\rm I}$ and the parameter $c_2$, while $M_{\rm G}$ shows a completely different behaviour. 
Since $a(r)\equiv 1$ due to the conformal invariance
the fact that the ratio $M_G/M_I$ is not constant clearly demonstrates that the equivalence principle does not 
hold in this case. 

In contrast to the case of AdS soliton solutions in Einstein-Hilbert gravity 
where two qualitatively different approaches to limiting solutions are available 
depending on the value of $e^2$ (see e.g. \cite{menagerie,bhs})
we find that here the qualitative pattern is independent of the choice of $e^2$.
This can be understood when remembering that in Einstein-Hilbert gravity and for small values of $e^2$ the solutions
tend to an ``attractor'' solution with the value of the metric function at the origin $a(0)\rightarrow 0$. 
In conformal gravity we can use the transformations to choose $a(r)\equiv 1$. So, it is natural that this limit does not exist
here.  

As an example let us describe the behaviour of
the gravitational mass $M_{\rm G}$ in more detail. For small values of $e^2$ (see Fig.\ref{small_e}) the solutions 
reach a maximum of the
gravitational mass $M_{\rm G}$ at an intermediate value of the charge $Q$, then 
the charge $Q$ becomes maximal at an intermediate value of the gravitational mass $M_{\rm G}$ and a second branch
of solutions extends backwards to smaller values of $Q$. Increasing $e^2$ further the curve nearly shows a spike-like behaviour
with the maximal value of $M_{\rm G}$ reached at the maximal value of $Q$. Increasing $e^2$ further, we observe that
a loop forms such that the second branch of solutions intersects the first branch at some value of $Q$. 
For even larger values of $e^2$ (see Fig.\ref{large_e}) we find that the solutions reach a solution with maximal
$Q$ at some intermediate $M_{\rm G}$ and then form a second branch of solutions extending back in $Q$ on which they
reach the solution with the largest $M_{\rm G}$. 

Finally, we have studied solutions with a fixed electric charge $Q$ for different values of the coupling $e^2$. 
We do not present a separate plot
here, because the qualitative features are the same as in the other cases. 
Let us simply mention that 
 the electric  potential $\phi(r)$ approaches  a constant in the limit $e^2 \to 0$. In this case, the U(1) gauge symmetry becomes
 global and hence the (originally) complex scalar field can not be gauge transformed to be real (as done throughout this paper).
 Hence, we would have to choose the scalar field complex and could e.g. adopt the following Ansatz with a periodic time-dependence
$\psi(r,t)=\tilde{\psi}(r) e^{i\omega t}$. 
Hence, $\phi(r) \equiv \mu$  can be reinterpreted as the frequency parameter $\omega$ characterizing the harmonic time-dependence of the soliton, which
makes the connection to boson star solutions apparent.

\section{Conclusion}
In this paper, we have studied the formation of conformal scalar hair on charged solitons in global (3+1)-dimensional AdS space-time.
Charged soliton solutions minimally coupled to gravity were studied in \cite{menagerie}. It was found that the pattern of solutions
is rich and crucially depends on the U(1) gauge coupling. In this paper, we have attempted to characterize the pattern of charged solitons 
(i) when the scalar field is non-minimally coupled to gravity and (ii) when considering conformal Weyl gravity instead of Einstein-Hilbert gravity, respectively.
Since we were aiming at studying  a matter action that is conformally invariant only a $\psi^4$-potential is compatible with this requirement. This excludes
the standard mass term for the scalar field which has been frequently studied in the context of AdS solitons with scalar hair. However, the conformal
coupling of the scalar field to gravity naturally induces a mass term for the scalar field in asymptotic AdS. 

One of the important results in the context of the self-interaction potential is that new branches of solutions exist that interpolate smoothly between 
the AdS vacuum and an uncharged configuration. A scaling argument indicates that these type of configurations should be possible
as soon as the self-coupling constant
of the scalar field is small enough. We find that we have to choose the self-coupling much smaller in the case of a minimally coupled scalar field 
as compared to the conformally coupled scalar field to find the interpolation between the AdS vacuum and a $Q=0$ configuration.
Furthermore, we find that in the case of conformal gravity the qualitative pattern of solutions seems to be independent of the gauge coupling
$e^2$ for the large range of the parameters that we have investigated. 

Consecutive to the evidence of particular instability channels of asymptotically AdS to formation of black holes \cite{Bizon:2011gg} a lot of
studies were devoted to this topic, in particular to the study of the stability of boson stars in asymptotically AdS (see e.g.
\cite{Dias:2012tq,Buchel:2012uh,Buchel:2013uba}). In particular, it was shown that AdS boson stars are non-pertubatively stable \cite{Dias:2012tq}.
Since our solitons have a boson star limit for vanishing gauge coupling and hence resemble these solutions it would be interesting to investigate
the different aspects of their stability in the context of conformal coupling as well as in conformal gravity.\\
\\
\\
\\
{\bf Acknowledgments} B.H. and S.T.
gratefully acknowledge support within the framework of the DFG Research
Training Group 1620 {\it Models of gravity}. Y.B. would like to thank the Belgian F.N.R.S. for financial
support.

\end{document}